\documentclass[%
10pt,
 oneocolumn,
 amsmath,amssymb,
 aps,
showkeys,
floatfix,
unsortedaddress
]{revtex4-2}

\usepackage[pdfpagelabels, pdfencoding=auto, psdextra]{hyperref}
\hypersetup{%
 pdfsubject=Paper,
 unicode = true,
 breaklinks = true,
 colorlinks = true,
 linkcolor = blue,
 citecolor = blue,
 menucolor = blue,
 citecolor = blue,
 urlcolor = blue
}

\usepackage{epsfig,graphicx,color}
\usepackage{booktabs}
\usepackage{float}   
\usepackage{graphicx}
\usepackage{subfig}
\usepackage{dcolumn}
\usepackage{bm}
\usepackage{multirow}
\usepackage{soul}

\newcommand{\nc}{\newcommand}

\nc{\half}{\frac{1}{2}}
\nc{\bp}{{\bf p}}
\nc{\bpp}{{\bf p'}}
\nc{\bpz}{{\bf p''}}
\nc{\bk}{{\bf k}}
\nc{\bkp}{{\bf k'}}
\nc{\bkz}{{\bf k''}}
\nc{\bra}{ \langle }
\nc{\ket}{ \rangle }
\nc{\bq}{{\bf q}}
\nc{\bqp}{{\bf q'}}
\nc{\tpi}{\tilde{\pi}}
\nc{\bpi}{\boldsymbol \pi}
\nc{\btpi}{\tilde{\boldsymbol \pi}}
\nc{\bu}{{\bf u}}
\nc{\bv}{{\bf v}}
\nc{\ba}{{\bf a}}
\nc{\bb}{{\bf b}}
\nc{\mK}{K}
\nc{\mH}{H}

\begin{document}

\title{Four-body bound states in momentum space: the Yakubovsky approach without two-body $t-$matrices}

\author{M. Mohammadzadeh}%
\affiliation{Department of Physics, K.~N.~Toosi University of Technology, P.O.Box 16315-1618, Tehran, Iran,}
\author{M. Radin}%
\affiliation{Department of Physics, K.~N.~Toosi University of Technology, P.O.Box 16315-1618, Tehran, Iran,}
\author{K.~Mohseni}%
\affiliation{Departamento de F\'isica, Instituto Tecnol\'ogico de Aeron\'autica, DCTA, 12228-900 S\~ao Jos\'e dos Campos, Brazil,}
\author{M.~R.~Hadizadeh}
\email[corresponding author:]{mhadizadeh@centralstate.edu} 
\affiliation{
College of Engineering, Science, Technology and Agriculture, Central State University, Wilberforce, OH
45384, USA, \\
Department of Physics and Astronomy, Ohio University, Athens, OH 45701, USA}%

\date{\today}

\begin{abstract}
This study presents a solution to the Yakubovsky equations for four-body bound states in momentum space, bypassing the common use of two-body $t-$matrices. Typically, such solutions are dependent on the fully-off-shell two-body $t-$matrices, which are obtained from the Lippmann-Schwinger integral equation for two-body subsystem energies controlled by the second and third Jacobi momenta. Instead, we use a version of the Yakubovsky equations that doesn't require $t-$matrices, facilitating the direct use of two-body interactions. This approach streamlines the programming and reduces computational time. 
Numerically, we found that this direct approach to the Yakubovsky equations, using 2B interactions, produces four-body binding energy results consistent with those obtained from the conventional $t-$matrix dependent Yakubovsky equations, for both separable (Yamaguchi and Gaussian) and non-separable (Malfliet-Tjon) interactions.

\end{abstract}

\maketitle

\section{Introduction}

The Yakubovsky equations provide a non-perturbative framework for investigating few-body bound and scattering systems in different sectors of physics, including atomic, nuclear, and particle physics. These equations have been utilized extensively in both momentum \citep{
kamada1992solutions,
glockle1993inclusion,
nogga2000modern,
nogga2002alpha,
uzu2003complex,
platter2004four,
nogga2005faddeev,
nogga2007application,
bayegan2007realistic,
deltuva2007four,
hadizadeh2007four,
bayegan2008three,
bayegan2008realistic,
hadizadeh2009towards,
hadizadeh2011solutions,
frederico2013universality,
tomio2013four,
kamada2019four,
hadizadeh2014relativistic,
hadizadeh2016relativistic,
radin2017four,
bazak2018energy,
bazak2019four,
hadizadeh2020three}
and configuration 
\citep{
barnea2000projected,
filikhin200016,
kezerashvili2001elastic,
filikhin20024he,
lazauskas2004ab,
uzu2006four,
lekala2010four,
schellingerhout2012numerical,
filikhin2014modeling,
lazauskas2015modern,
hiyama2016possibility,
kezerashvili2017few,
kezerashvili2017strange,
ferrari2017benchmark,
lazauskas2018solution,
lazauskas2019faddeev,
lazauskas2019parity,
kezerashvili2019few,
filikhin2020three,
lazauskas2020description,
kezerashvili2021search} spaces.
The Yakubovsky equations fundamentally rely on two-body (2B) transition matrices, denoted as $t(\epsilon)$, which are derived from the solution of the Lippmann-Schwinger (LS) equation, considering either positive (scattering states) or negative (bound states) 2B subsystem energies $\epsilon$. Notably, solving the LS equation for positive energies can prove to be a numerically demanding task due to the presence of singularities. Contrarily, for negative energies, the LS equation must be computed for the 2B subsystem energies, which are determined by the Jacobi momenta of the third and fourth particles.

In this study, we utilize a version of the Yakubovsky equations for four-body (4B) bound states that directly incorporates 2B interactions, eliminating the need for 2B $t-$matrices. This $t-$matrix-free approach to the Yakubovsky equations has been previously solved in configuration space by Lazauskas {\it et al.} \cite{lazauskas2004ab, lazauskas2018solution, lazauskas2019faddeev, lazauskas2020description}. In our work, we present the $t-$matrix-free coupled Yakubovsky equations in momentum space. Here, the Yakubovsky components linked to the $3+1$ and $2+2$ chains are derived as a function of the Jacobi momentum vectors, directly including 2B interactions. To determine the 4B binding energies, we solve the coupled Yakubovsky integral equations using separable potentials with Yamaguchi and Gaussian form factors, as well as the non-separable Malfliet-Tjon potential, with all potentials projected into the $s-$wave channel. 
Our numerical findings highlight that the $t-$matrix-free version of the Yakubovsky integral equations, when utilizing 2B interactions, aligns perfectly with results derived from the conventional Yakubovsky integral equations that employ 2B $t-$matrices. 
In a related research, this $t-$matrix-free approach is successfully employed for relativistic three-body (3B) bound states \citep{Mohammadzadeh_submitted}. This led to a version of the relativistic Faddeev equation that directly employs 2B boosted interactions \citep{hadizadeh2017calculation, hadizadeh2021relativistic}, eliminating the need for 2B boosted $t-$matrices \citep{hadizadeh2014relativistic, hadizadeh2016relativistic, hadizadeh2020three}.

This paper has been structured into several sections. Section \ref{sec.Yakubovsky} provides a brief overview of the Yakubovsky equations for 4B bound states, bypassing the use of 2B $t-$matrices. Herein, we formulate the coupled Yakubovsky equations in momentum space, considering Jacobi momentum vectors and presenting a simplified form for $s-$wave interactions. Section \ref{sec.results} outlines our numerical results for 4B binding energies as calculated from the $t-$matrix-free Yakubovsky equations, alongside a comparison with results obtained from conventional $t-$matrix-dependent Yakubovsky equations. Lastly, Section \ref{sec.conclusion} presents our conclusion and discusses prospects for succeeding studies.

\section{The $t-$matrix-free coupled Yakubovsky equations for 4B bound states}\label{sec.Yakubovsky}
The conventional form of Yakubovsky equations that describe the bound state of four identical particles interacting through pairwise interactions reads as follows \citep{nogga2002alpha}
\begin{eqnarray}\label{eq_Yakubovsky_t_operator}
	\psi_1   &=&  G_0 t  P \left [ (1+P_{34})\psi_1 +\psi_2 \right ], \cr
	\psi_2  &=&  G_0 t  \tilde{P}  \left [ (1+P_{34})\psi_1 + \psi_2 \right],
\end{eqnarray}
where $\psi_1$ and $\psi_2$ denote Yakubovsky components of the 4B wave function, corresponding to $3+1$ and $2+2$ type chains, respectively. $G_0=(E-H_0)^{-1}$ represents 4B free propagator, while $P=P_{12}P_{23}+P_{13}P_{23}$, $\tilde{P}=P_{13}P_{24}$ and $P_{34}$ serve as the permutation operators. The 2B $t-$matrix is determined by the LS equation
\begin{equation} \label{t-matrix}
t = V + V G_0 t.
\end{equation}
The coupled Yakubovsky equations \eqref{eq_Yakubovsky_t_operator} can be restructured to yield another form of Yakubovsky equations as follows
\begin{eqnarray}\label{eq_Yakubovsky_v_operator}
\psi_1  &=&  G_0VP \left [ (1+P_{34})\psi_1 +\psi_2 \right ] + G_0V\psi_1, \cr
\psi_2  &=&  G_0V\tilde{P} \left [ (1+P_{34})\psi_1 +\psi_2 \right ] + G_0V\psi_2,
\end{eqnarray}
where the 2B interactions $V$ are being directly utilized as input to the Yakubovsky equations, consequently eliminating the need for the 2B $t-$matrices.
The representation of the $t-$matrix-free form of the coupled Yakubovsky equations \eqref{eq_Yakubovsky_v_operator} in momentum space leads to the following coupled 3D integral equations
\begin{eqnarray}
  \psi_1 (\bu_1 , \bu_2 , \bu_3)  
 &=&
G_0(u_1,u_2,u_3) 
\Biggl (
 \int d^{3} u'_2  
 V_{s} \left (\bu_1, \frac{1}{2}\bu_2 + \bu'_2 \right) 
\cr 
&& \hskip-2cm  \times
\biggl \{
\psi_1 \left (\bu_2 +\frac{1}{2} \bu'_2, \bu'_2, \bu_3 \right )
 + 
 \psi_1 \left(  \bu_2 +\frac{1}{2} \bu'_2, \frac{1}{3}\bu'_2 + \frac{8}{9}\bu_3,
\bu'_2-\frac{1}{3}\bu_3  \right)
 +
 \psi_2 \left(  \bu_2 +\frac{1}{2} \bu'_2, -\bu'_2-\frac{2}{3}\bu_3, \frac{1}{2}\bu'_2-\frac{2}{3}\bu_3 \right)
  \biggr \} 
  \cr
  & &  \hskip-2cm + 
  \frac{1}{2}  \int  d^{3}u'_1 
 V_{s} \left( \bu_1, \bu'_1 \right)
 \psi_1 \left( \bu'_1, \bu_2, \bu_3 \right)
  \Biggr ),
\cr\cr
 \bra  \bv_1, \bv_2,\bv_3|\psi_2 \ket  &=& 
\frac{1}{2}G_0(v_1,v_2,v_3)
\Biggl (
 \int  d^{3} v_3' 
 V_{s} \left(\bv_1,  \bv'_3 \right)
\cr
&& \hskip-2cm \times
\biggl \{ 
2  
\psi_1 \left( \bv_3, \frac{2}{3}\bv_2+\frac{2}{3}\bv'_3, \frac{1}{2}\bv_2-\bv'_3  \right)
 +
 \psi_2 \left(  \bv_3, -\bv_2, \bv'_3 \right)
 \biggr \}
+   \int  d^{3}v'_1  
V_{s} \left(\bv_1, \bv'_1 \right)
\psi_1  \left( \bv'_1,  \bv_2,\bv_3 \right)
    \Biggr ) ,
\label{eq.Yakubovsky_v_momentum}
 \end{eqnarray}
with the symmetrized 2B interaction defined as $V_{s} (\ba, \bb) = V (\ba, \bb) + V (\ba, -\bb)$ and 4B free propagators characterized through the following expressions
\begin{equation}\label{eq.G0_momentum}
G_0(u_1,u_2,u_3)  = \bigg(E-\frac{u_1^2}{m}-\frac{3u_2^2}{4m}-\frac{2u_3^2}{3m} \bigg)^{-1}, \quad
G_0(v_1,v_2,v_3) = \bigg(E-\frac{v_1^2}{m}-\frac{v_2^2}{2m}-\frac{v_3^2}{m}\bigg)^{-1}.
\end{equation}
The Jacobi momenta $\bu_i$ and $\bv_i$ ($i=1,2, 3$) correspond to $3+1$ and $2+2$ chains, respectively \citep{hadizadeh2007four}. 
The coupled Yakubovsky equations \eqref{eq.Yakubovsky_v_momentum} can be simplified for the $s-$wave interactions as
\begin{eqnarray} \label{eq.nonSep_V_YC}
  \psi_1 (u_1, u_2, u_3 ) &=& 
  4 \pi \, G_{0} ( u_1, u_2, u_3)
  \cr
    &\times&
  \Biggl (
   \,  \int_0^\infty  du'_2 u'^2_2 \int_{-1}^1  dx' \
  V \bigl (u_1,\Pi_1 (u_2,u'_2,x' )  \bigr)
  \cr
  &\times&
  \Biggl[
  \psi_1 \bigl (\Pi_1 (u'_2,u_2,x' ) , u'_2, u_3  \bigr) 
  + \frac{1}{2}\int_{-1}^1 dx\,
  \Biggl\{ 
  \psi_1 \bigl (\Pi_1 (u'_2,u_2,x' ) , \Pi_2 (u'_2,u_3,x ), \Pi_3 (u'_2,u_3,x  )  \bigr) 
  \cr &+&
  \psi_2  \bigl (\Pi_1  (u'_2,u_2,x'  ) , \Pi_4 (u'_2,u_3,x  ), \Pi_5 (u'_2,u_3,x )  \bigr)
  \Biggr\} 
  \Biggr]
  +  \int_0^\infty d u'_1 u'^2_1 \, V \bigl(u_1,u'_1 \bigr) \, 
  \psi_1 \bigl(u'_1, u_2, u_3 \bigr)   \Biggr ), 
  \cr
  \psi_2  \bigl(v_1, v_2, v_3  \bigr) &=& 4 \pi \, G_{0}  ( v_1, v_2, v_3 ) 
  \cr
  &\times&  \Biggl (
   \int_0^\infty
  dv'_3 v'^2_3 \, V  \bigl(v_1,v'_3 \bigr)
  \Biggl[\int_{-1}^1  dx'_3\, 
  \psi_1  \bigl(v_3, \Pi_6(v_2,v'_3,x'_3  ), \Pi_7  (v_2,v'_3,x  )   \bigr)
   + \psi_2  \bigl(v_3, v_2, v'_3  \bigr) 
   \Biggr]
   \cr
   && +    \int_0^\infty d v'_1 v'^2_1 \, 
   V \bigl(v_1,v'_1 \bigr)  \,\psi_2 \bigl(v'_1, v_2, v_3 \bigr)   
   \Biggr ),
   \end{eqnarray}
where the shifted momentum arguments are defined as \citep{hadizadeh2012binding}
\begin{eqnarray}  \label{eq.shifted_momenta}
&& \tilde{\Pi}_1 = \left(\frac{1}{4}u_2^2+u'^{2}_2+u_2u'_2  x \right)^{\frac{1}{2}},
\quad 
\Pi_1 = \left(u_2^2+\frac{1}{4}u'^{2}_2+u_2u'_2  x \right)^{\frac{1}{2}},
\cr
&& \Pi_2 = \frac{1}{3} \left(u_2^2+\frac{64}{9}u_3^{2}+\frac{16}{3}u_2u_3  x' \right)^{\frac{1}{2}},
\quad
\Pi_3 = \left(u_2^2+\frac{1}{9}u_3^{2}-\frac{2}{3}u_2u_3  x' \right)^{\frac{1}{2}},
\cr
 && \Pi_4 = \left(\frac{1}{4}u_2^2+\frac{4}{9}u_3^{2}+\frac{2}{3}u_2u_3  x' \right)^{\frac{1}{2}},
\quad
\Pi_5 = \left(u_2^2+\frac{4}{9}u_3^{2}-\frac{4}{3}u_2u_3  x' \right)^{\frac{1}{2}},
\cr
&& \Pi_6 = \frac{2}{3} \left(v_2^2+v_3^{2}+2v_2v_3  x \right)^{\frac{1}{2}},
\quad
\Pi_7 = \left(v_3^2+\frac{1}{4}v_2^{2}-v_2v_3  x \right)^{\frac{1}{2}}.
\end{eqnarray}
For comparison purposes, the representation of the conventional $t-$matrix-dependent form of the coupled Yakubovsky equations \eqref{eq_Yakubovsky_t_operator} in momentum space can be found in Appendix \ref{Appendix_Yakubovsky_t}. Upon comparing the $t-$matrix-free and conventional $t-$matrix-dependent forms of the Yakubovsky equations - specifically, Eqs. \eqref{eq.Yakubovsky_v_momentum} and \eqref{eq.Yakubovsky_t_momentum} - it is evident that the $t-$matrix-free form incorporates an extra term. This term involves integration over the 2B interaction and the Yakubovsky components without interpolations on momenta or angles. Despite this, its numerical solution proves to be more straightforward and cost-effective than that of the conventional $t-$matrix-dependent that necessitates solving the LS equation to compute the 2B $t-$matrices for all required 2B subsystem energies, which depend on the magnitude of the second and third Jacobi momenta.

\section{Numerical results}\label{sec.results}

The numerical solution of the coupled Yakubovsky integral equations \eqref{eq.nonSep_V_YC} for the calculation of 4B binding energy demands solving an eigenvalue equation, where the physical binding energy corresponds to an eigenvalue equal to one. The Lanczos iterative method is implemented for solving such eigenvalue equation \citep{hadizadeh2012binding,hadizadeh2020three,ahmadi2020novel,mohseni2021three,hadizadeh2011scaling}.
We employ Gauss-Legendre quadratures to discretize the continuous momentum and angular variables with a hyperbolic mapping for Jacobi momenta and a linear mapping for angle variables \citep{mohseni2023trion}. This allows us to properly capture the behavior of the Yakubovsky components of the 4B wave function at both small and large momenta. 
In each iteration step of solving the coupled Yakubovsky integral equations, to accurately perform multi-dimensional interpolations on the shifted momentum arguments given in Eq.~\eqref{eq.shifted_momenta}, we employ the Cubic-Hermite spline method due to its combination of high accuracy and computational speed \citep{huber1997new}.

Our numerical analysis presents a comparison between 4B binding energy obtained from the $t-$matrix-free approach and the conventional $t-$matrix-dependent formulation of the coupled Yakubovsky equations, namely, equations \eqref{eq.nonSep_V_YC} and \eqref{Yakubovsky_t_s_wave}. For our numerical analysis, we employ two models of one-term separable potential with the following general form
\begin{equation}
V(p,p')= \lambda g(p) g(p'),
\end{equation}
where $\lambda$ represents the potential strength. The potential form factor $g(p)$ for the Yamaguchi-type potential is defined as $g(p)=1/(p^2+\beta^2)$ \citep{PhysRev.95.1628}, while for the Gaussian potential, it takes the form $g(p)=\text{exp}(-p^2/\Lambda^2)$ \citep{deltuva2011universality}.
Furthermore, to provide a comprehensive validation of our formalism and code, we also incorporate an $s$-wave non-separable Malfliet-Tjon (MT) potential, comprises two attractive and repulsive terms \cite{kessler2003scattering}
\begin{equation}
  V(p, p') = \sum_{i=1}^2 \frac{\lambda_i}{2 \pi p p'} \ln \left(\frac{\mu_i^2+p^2+p'^2+2 p p'}{\mu_i^2+p^2+p'^2-2 p p'}\right).
\end{equation}
In Table \ref{potentials_parameter}, we provide the parameters for the potentials employed in our calculations, which include MT model V (MT-V), Yamaguchi potential model IV (Y-IV), and a Gaussian potential. The strength of the Gaussian potential was adjusted to reproduce the deuteron binding energy of $-2.225$ MeV, with a form factor parameter $\Lambda = 0.7 \ \text{fm}^{-1}$.
\begin{table}[hbt]
  \centering
  \caption{The parameters for the Malfliet-Tjon (MT-V), Yamaguchi (Y-IV), and Gaussian potentials utilized in this study.}
  \label{potentials_parameter}
  \begin{tabular}{cc|cccc}
      \toprule
  MT-V  &&    $\lambda_1$ &  $\mu_1$ (fm$^{-1}$) & $\lambda_2$ & $ \mu_2$ (fm$^{-1}$) \\
      \midrule
    &&  $-2.93$ & $1.55$ & $7.39$ & $3.11$  \\
      \midrule
  Yamaguchi-IV &&   $\lambda$ (MeV$\cdot$fm$^{-1}$) &  $\beta$ (fm$^{-1}$) && \\
      \midrule
    &&  $-7.42313$  & $1.15$ && \\
      \midrule
  Gaussian  &&   $\lambda$ (MeV$\cdot$fm$^{3}$)  &  $\Lambda$ (fm$^{-1}$) &&  \\
      \midrule
    && $-15.825$  &  $0.70$ && \\
      \bottomrule
  \end{tabular}
\end{table}

Table \ref{table:4B_energiy_convergence} presents the convergence of the 4B binding energy as a function of the number of mesh points for the Jacobi momenta magnitudes $u_i$ and $v_i$. The table provides a side-by-side comparison of results obtained using both the $t$-matrix-free and conventional $t$-matrix-dependent forms of the coupled Yakubovsky equations. Specifically, our results show that the $t-$matrix-free Yakubovsky equations yield a 4B binding energy of $-30.08$ MeV for the MT-V potential and $-36.27$ MeV for the Yamaguchi-IV potential. These values closely align with the $-30.07$ MeV \cite{KAMADA1992205}  and $-36.3$ MeV \cite{PhysRevC.15.2257.2} obtained by other groups using the $t-$matrix-dependent Yakubovsky calculations. The comparison validates the potential of the $t-$matrix-free adoption of Yakubovsky formulation to solve 4B bound state problems efficiently, matching the precision of the conventional method but potentially offering more straightforward computational requirements. Moreover, the convergence behavior remains consistent across different numbers of mesh points for the magnitude of Jacobi momenta, emphasizing the computational robustness of the $t-$matrix-free formulation.

\begin{table}[!htb]
\centering
\caption{Convergence of the 4B binding energy obtained from the $t-$matrix-free ($E_v$), Eq. \eqref{eq.nonSep_V_YC}, and the conventional $t-$matrix-dependent ($E_{\text{t-matrix}}$), Eq. \eqref{Yakubovsky_t_s_wave}, versions of the coupled Yakubovsky equations. The convergence is shown as a function of the number of mesh points for the magnitude of Jacobi momenta $u_i$ and $v_i$, denoted as $N_{u_i} = N_{v_i}$. Results are presented for MT-V, Yamaguchi-IV, and Gaussian potentials. The number of mesh points for angle variables is 40. All calculations were performed with $\hbar^2/m = 41.47$ MeV $\cdot$ fm$^2$.}
\label{table:4B_energiy_convergence}
\begin{tabular}{c|cc|cc|cc}
$N_{u_i} = N_{v_i}$  & $E_{\text{v}}$ (MeV) & $E_{\text{t-matrix}}$ (MeV) & $E_{\text{v}}$ (MeV) & $E_{\text{t-matrix}}$ (MeV)& $E_{\text{v}}$ (MeV) & $E_{\text{t-matrix}}$ (MeV)
\\   \toprule
&  \multicolumn{2}{c}{$s-$wave MT-V} &\multicolumn{2}{c}{Yamaguchi-IV} & \multicolumn{2}{c}{Gaussian} \\  
\cline{2-3}\cline{4-5}\cline{6-7}
$30$ & $-30.323$ & $-30.668$ &  $-36.156$ &  $-36.151$ & $-31.281$ & $-31.276$  \\
$40$ & $-30.206$ & $-30.367$ &  $-36.236$ &  $-36.233$ & $-31.272$ & $-31.272$ \\
$50$ & $-30.313$ & $-30.220$ &  $-36.258$ &  $-36.256$ & $-31.272$ & $-31.271$ \\
$60$ & $-30.182$ & $-30.145$ &  $-36.266$ &  $-36.264$ & $-31.271$ & $-31.271$ \\
$70$ & $-30.135$ & $-30.116$ &  $-36.270$ &  $-36.268$ & $-31.271$ & $-31.271$ \\
$80$ & $-30.103$ & $-30.098$ &  $-36.272$ &  $-36.270$ & $-31.271$ & $-31.270$ \\
$90$ & $-30.092$ & $-30.089$ &  $-36.273$ &  $-36.271$ & $-31.270$ & $-31.270$ \\
$100$ & $-30.087$ & $-30.084$ & $-36.273$ &  $-36.271$ & $-31.270$ & $-31.270$ \\
$150$ & $-30.078$ & $-30.077$ & $-36.274$ &  $-36.272$ & $-31.270$ & $-31.270$ \\
\hline
 &  $-$ & $-30.07$ \cite{KAMADA1992205} & $-$ &  $-36.3$ \cite{PhysRevC.15.2257.2} & $-$ & $-$ \\
\bottomrule
\end{tabular}
\end{table}

%
\section{Summary and Outlook}\label{sec.conclusion}
The Yakubovsky approach is a powerful method to study few-body bound and scattering systems. However, the solution of these equations can be computationally demanding due to inherent singularities in 2B $t-$matrices when dealing with scattering problems and the need to calculate them for 2B subsystem energies dictated by second and third Jacobi momenta when 4B bound state problems are considered. This study utilizes a version of the coupled Yakubovsky equations for 4B bound states that directly incorporates 2B interactions in momentum space, avoiding the use of the 2B $t-$matrices. The efficacy of this approach is validated through the calculation of 4B binding energies in momentum space using, both the separable potentials with Yamaguchi and Gaussian form factors, and the non-separable Malfliet-Tjon potential. Our findings align well with results from the conventional form of the coupled Yakubovsky integral equations incorporating 2B $t-$matrices. The extension of calculations to include more general interactions, beyond just the $s-$wave, is currently in progress.

\appendix

\section{Representation of conventional $t-$matrix-dependent coupled Yakubovsky equations in momentum space} \label{Appendix_Yakubovsky_t}
The conventional form of the coupled Yakubovsky equations \eqref{eq_Yakubovsky_t_operator} in momentum space is represented as follows \citep{hadizadeh2007four}
\begin{eqnarray}
\psi_1 \left( \bu_1 , \bu_2 , \bu_3 \right)
 &=&
G_0(u_1,u_2,u_3) \int d^{3} u'_2  
 \ t_{s} \left (\bu_1, \frac{1}{2}\bu_2 + \bu'_2 ;\epsilon \right)
\cr 
&\times& 
\biggl \{
\psi_1 \left( \bu_2 +\frac{1}{2} \bu'_2, \bu'_2, \bu_3 \right)
 + 
 \psi_1 \left( \bu_2 +\frac{1}{2} \bu'_2, \frac{1}{3}\bu'_2 + \frac{8}{9}\bu_3, \bu'_2-\frac{1}{3}\bu_3 \right)
 \cr  
 && +
 \psi_2  \left( \bu_2 +\frac{1}{2} \bu'_2, -\bu'_2-\frac{2}{3}\bu_3, \frac{1}{2}\bu'_2-\frac{2}{3}\bu_3 \right)  
  \biggr \},
\cr\cr
\psi_2  \left(   \bv_1, \bv_2,\bv_3 \right)
 &=& \frac{1}{2}G_0(v_1,v_2,v_3) \int  d^{3}v_3' 
 \ t_{s} \left( \bv_1,  \bv'_3 ; \epsilon^{*} \right)
\cr  &\times&  
\biggl \{ 
2  \psi_1 \left( \bv_3, \frac{2}{3}\bv_2+\frac{2}{3}\bv'_3, \frac{1}{2}\bv_2-\bv'_3 \right)
 +
 \psi_2 \left( \bv_3, -\bv_2, \bv'_3 \right)
 \biggr \} ,
\label{eq.Yakubovsky_t_momentum}
 \end{eqnarray}
with the symmetrized 2B $t-$matrices defined as
\begin{equation}
t_{s}(\ba,  \bb ; \epsilon ) = t(\ba,  \bb ; \epsilon ) + t(\ba,  -\bb ; \epsilon ) .
\end{equation}
The matrix elements of 2B $t-$matrices $t(\ba,  \bb ; \epsilon )$ needs to be calculated from the solution of the LS equation \eqref{t-matrix} for the 2B subsystem energies associated with the $3+1$ and $2+2$ chains, given by 
\begin{eqnarray}
\epsilon  = E-\frac{3u_2^2}{4m}-\frac{2u_3^2}{3m} , \quad\quad\quad
\epsilon^* = E-\frac{v_2^2}{2m}-\frac{v_3^2}{m}.
\end{eqnarray}
The coupled Yakubovsky equations \eqref{eq.Yakubovsky_t_momentum} can be simplified for the $s-$wave interactions as
\begin{eqnarray} \label{Yakubovsky_t_s_wave}
  \psi_1 \bigl(u_1, u_2, u_3 \bigr) &=& 4 \pi \, G_{0} \bigl( u_1, u_2,
  u_3\bigr) \, 
   \int_0^\infty  du'_2 u'^2_2 \int_{-1}^1  dx' \
  t \bigl(u_1,\Pi_1\bigl(u_2,u'_2,x' \bigr), \epsilon \bigr)\
  \cr
  &\times&
  \Biggl(
  \psi_1 \bigl(\Pi_1 \bigl(u'_2,u_2,x' \bigr) , u'_2, u_3 \bigr)
  + \frac{1}{2}\int_{-1}^1 dx\,
  \Biggl\{ 
  \psi_1 \bigl(\Pi_1 (u'_2,u_2,x') , \Pi_2(u'_2,u_3,x), \Pi_3(u'_2,u_3,x) \bigr)
  \cr && +
  \psi_2 \bigl(\Pi_1 (u'_2,u_2,x') , \Pi_4 (u'_2,u_3,x), \Pi_5(u'_2,u_3,x) \bigr)
  \Biggr\} 
  \Biggr), 
  \cr
  \psi_2 \bigl(v_1, v_2, v_3 \bigr) &=& 4 \pi \, G_{0} ( v_1, v_2,
  v_3) \, 
 \int_0^\infty
  dv'_3 v'^2_3 \, t (v_1,v'_3,\epsilon^*)
    \cr
  &\times&
  \Biggl(
  \int_{-1}^1  dx'_3\, 
  \psi_1 \bigl(v_3, \Pi_6(v_2,v'_3,x'_3), \Pi_7 (v_2,v'_3,x)  \bigr)
   + \psi_2 (v_3, v_2, v'_3) 
   \Biggr).
   \end{eqnarray}
\section*{Conflict of Interest Statement}
The authors declare that the research was conducted in the absence of any commercial or financial relationships that could be construed as a potential conflict of interest.

\section*{Author Contributions}
M.R. took the lead in project design, while M.R. and M.R.H. collaborated on developing the theoretical formalism and computer codes. M.M. and K.M. calculated the four-body binding energies. M.R.H., M.R., and K.M. engaged in the result discussion and contributed to the final manuscript.

\section*{Funding}
The work of M.R.H. was supported by the National Science Foundation under Grant No. NSF-PHY-2000029 with Central State University. K.M. was supported by Conselho Nacional de Desenvolvimento Cient\'ifico e Tecnol\'ogico (CNPq) Grant No.~400789/2019-0.

\section*{Acknowledgments}
K.M. acknowledges a Ph.D. scholarship from the Brazilian agency CNPq. 

\bibliography{References}

\end{document}